\documentclass[traditabstract]{aa}

\usepackage{graphics}   % standard graphics specifications
\usepackage{graphicx}   % alternative graphics specificati.jpgons
\usepackage{epstopdf}
\usepackage{longtable}   % helps with long table options
\usepackage{url}      % for on-line citations
\usepackage{bm}        % special 'bold-math' package
\usepackage{longtable}   % helps with long table options

\usepackage[varg]{txfonts}

\usepackage{supertabular}
\usepackage{hyperref}
\usepackage{amsmath}
\usepackage[table,svgnames]{xcolor}
\usepackage{booktabs}

\usepackage{pdflscape}
\usepackage{supertabular}

\definecolor{green}{rgb}{0.3,0.7,0.}

\hypersetup{
    bookmarks=true,         % show bookmarks bar?
    unicode=false,          % non-Latin characters in Acrobat?s bookmarks
    colorlinks=true,       % false: boxed links; true: colored links
    linkcolor=blue,          % color of internal links (change box color with linkbordercolor)
    citecolor=blue,        % color of links to bibliography
    filecolor=blue,      % color of file links
    urlcolor=blue           % color of external links
}
\usepackage{color}

\newcommand{\msun}{\ensuremath{M_{\odot}}}

\newcommand\Mpy{\ensuremath{\msun\,{\rm yr}^{-1}}}
\newcommand\gva{{\sc Genec}}

\begin{document}
   \title{Linear adiabatic analysis for general relativistic instability in primordial accreting supermassive stars}

   % \subtitle{}

   \author{Hideyuki Saio\inst{1} 
   \and Devesh Nandal\inst{2}
  \and Sylvia Ekstr\"om\inst{2}
  \and George Meynet\inst{2}
  }
   
   \authorrunning{Saio et al.}

   \institute{Astronomical Institute, Graduate School of Science, Tohoku University,  Sendai, 980-8578, Japan\\
   \email{saio@astr.tohoku.ac.jp}
   \and D\'epartement d'Astronomie, Universit\'e de Gen\`eve, Chemin Pegasi 51, CH-1290 Versoix, Switzerland %}
   \\
   \email{Devesh.Nandal@unige.ch,
   sylvia.ekstrom@unige.ch, Georges.Meynet@unige.ch}
}

%   \date{Received ?; accepted ?}
   
  \abstract
  {Accreting supermassive stars of $\gtrsim 10^{5}\,\msun$ will eventually collapse directly to a black hole via the general relativistic (GR) instability.
  Such direct collapses of supermassive stars are thought to be a possible formation channel for supermassive black holes at $z > 6$.
  In this work, we investigate the final mass of accreting Population III stars with constant accretion rates between $0.01$ and 1000\,\Mpy.
 We determine the final mass by solving the differential equation for the general relativistic linear adiabatic radial pulsations. We find that models with accretion rates $\gtrsim 0.05\Mpy$ experience the GR instability at masses depending on the accretion rates. The critical masses are  larger for higher accretion rates, ranging from  $8\times10^{4}\,\msun$ for $0.05\,\Mpy$ to $\sim10^6\,$\msun\, for $1000\,\Mpy$.
  The $0.05\,\Mpy$ model reaches the GR instability at the end of the core hydrogen burning. 
  The higher mass models with the higher accretion rates reach the GR instability during the hydrogen burning stage.
}
   \keywords{Instabilities,Stars:formation,Stars:massive,Stars:evolution,Stars:Population III}

   \maketitle
%
%________________________________________________________________

\section{Introduction}
The observation of supermassive black holes with masses exceeding $10^{9}$ M$_\odot$ at redshifts $z > 6$ has prompted an exploration into their formation process \citep[e.g.][]{rees1978,Mortlock2011,woods2019}.
Possible building blocks of the super massive black holes are primordial (Pop\,III) supermassive stars (SMSs) of $10^4-10^6\,\msun$ which can be formed by very rapid accretion rates of $\dot{M} \gtrsim 0.1\,\Mpy$ \citep[e.g.][Nandal et al. 2024 submitted to A\&A]{Hosokawa2013,woods2017,Lionel2018,Herrington2023}. %\textcolor{green}
 SMS formation with even higher accretion rates has been proposed by \citet{Mayer2010, Mayer2019, Zwick2023}, where they show that early galaxy collisions can torque gas at up to solar metallicities into supermassive nuclear disks (SNDs). These SNDs may experience radial gravitational instabilities, collapsing at rates up to $10^5\,\msun$ yr$^{-1}$. The central object formed under such extreme conditions is expected to be short-lived and undergo "dark collapse" at masses above $10^6\,\msun$ \citep{Lionel2020, Lionel2021}.

The accreting object can become so massive that it encounters the general-relativistic (GR) instability \citep{Iben1963,chandrasekhar1964,fowler1966}. For masses below 10$^{5}$ M$_\odot$, the GR instability is unlikely to be encountered during the core hydrogen burning phase \citep{woods2019}. 
However, it may be reached in later evolutionary stages via the pair-instability process \citep{umeda2016,woods2017,woods2020}. Collapse of such a star with mass below $10^5\,\msun$ may result in an explosion, neutrino emissions, and an ultra long gamma ray burst, as per studies by \cite{Chen2014} and \cite{Nagele2021, Nagele2022}. 
For Pop III stars exceeding $10^5\,\msun$, the GR instability is encountered either before or during the core hydrogen-burning phase. Once initiated by the GR instability, the collapse proceeds unimpeded, resulting in the star's direct transition into a black hole without undergoing an explosion, thereby avoiding mass loss \citep{Fuller1986,montero2012}.
 
Upper mass limits of SMSs due to the general relativistic instability (GRI) have been studied by different methods.
\citet{Iben1963} found the binding energy of static SMSs to decrease with mass and to become 
negative at a mass between $10^5$ and $10^6\msun$.
From the mass dependence of the total energy, \citet{Osaki1966} found that the SMSs more massive than $3.5\times10^5\msun$ encounter the GRI before the onset of hydrogen burning.
\citet{umeda2016} determined the GRI critical mass of an accreting SMS at a rapid hydrostatic contraction of the post Newtonian hydrostatic structure.
On the other hand, using evolution/hydrodynamic codes with post Newtonian corrections, \citet{woods2017} and \citet{Herrington2023} detected the occurrence of GRI as hydrodynamic collapses of their models. 
The critical masses obtained from hydrostatic calculations are systematically larger than those from the  hydrodynamic calculations \citep[see Fig.\,11 in][]{Herrington2023}.
The latter seem appropriate, because the hydrostatic evolution code tends to average out small changes on short timescales that lead to large changes in the unstable structure.
For this reason, GR hydrodynamics are needed to  correctly detect the encounter with the GRI during the SMS evolution. 

Another way to find the occurrence of the GRI is based on the linear perturbation analysis, in which the small structure change is expressed as the consequence of a small radial displacement $\xi(r)\,e^{i\sigma ct}$ from the equilibrium position. 
The displacement $\xi$ is governed by a homogeneous differential equation with an eigenvalue $\sigma^2$ \citep[][]{chandrasekhar1964} (eq. (\ref{eq:ch64}) in \S\ref{puls} below). 
We see the occurence of the GRI for a star if we solve the differential equation for the stellar structure and find $\sigma^2 < 0$.
Solving the eigenvalue problem, \citet{Nagele2022} found that stars with $2 - 4\times10^4M_\odot$ (without accretion) become GR unstable during or after the core helium burning stage. 
On the other hand, assuming $\xi/r$ to be constant in the stellar interior,
\citet{Lionel2021} 
derived an integral expression (which consists of only structure variables) 
for $\sigma^2$ to judge the GR instability in  stellar evolution models.  
\citeauthor{Lionel2021} derived critical masses of $2.29$ and $4.37\times10^5M_\odot$ for accretion rates of 1 and $10\Mpy$, respectively. 

In this paper, we study the 
GRI in rapidly accreting Pop III SMS  models obtained by Nandal, et al.\,(2024 submitted A\&A) under the post Newtonian (PN) gravity. For these models we solve adiabatic linear pulsation equation of \citet{chandrasekhar1964} to obtain $\xi$ and $\sigma^2$. 

In Section\,\ref{sec-models} we show briefly the properties of the Pop III accreting SMS models to which our stability analysis is applied. 
In Section\,\ref{puls} we discuss the equations employed in our stability analysis. We present the results in Section\,\ref{sec:stabil} and compare them with previous results in Section\,\ref{sec:comparison}. 
In Appendix, we discuss briefly the numerical stability analysis applied to the n=3 PN polytrope to test our PN pulsation code.

\section{Primordial (Pop III) stellar evolution models with rapid mass accretion 
\label{sec-models}}

\begin{figure*}
    \centering
    \includegraphics[width=0.48\textwidth]{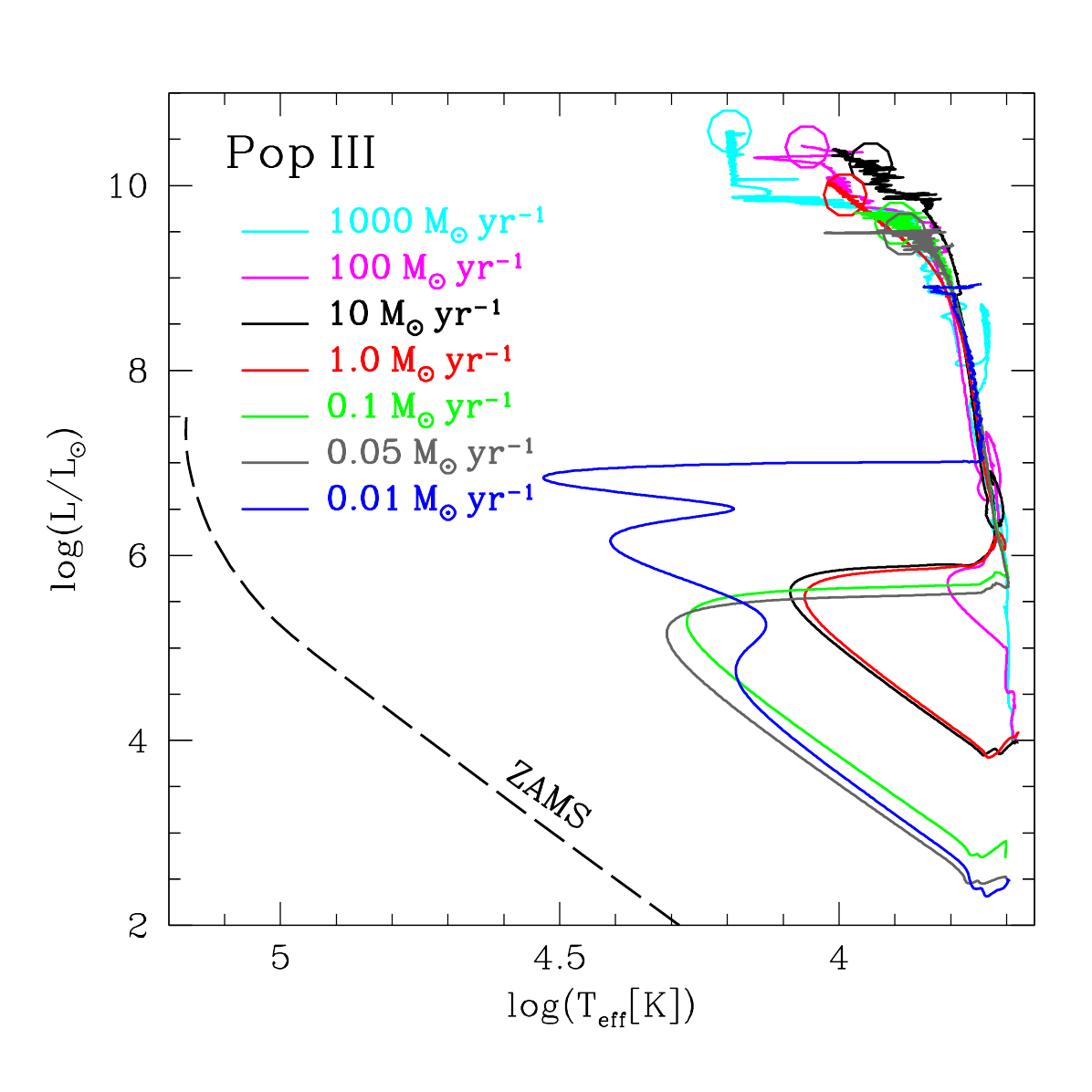} 
    \includegraphics[width=0.48\textwidth]{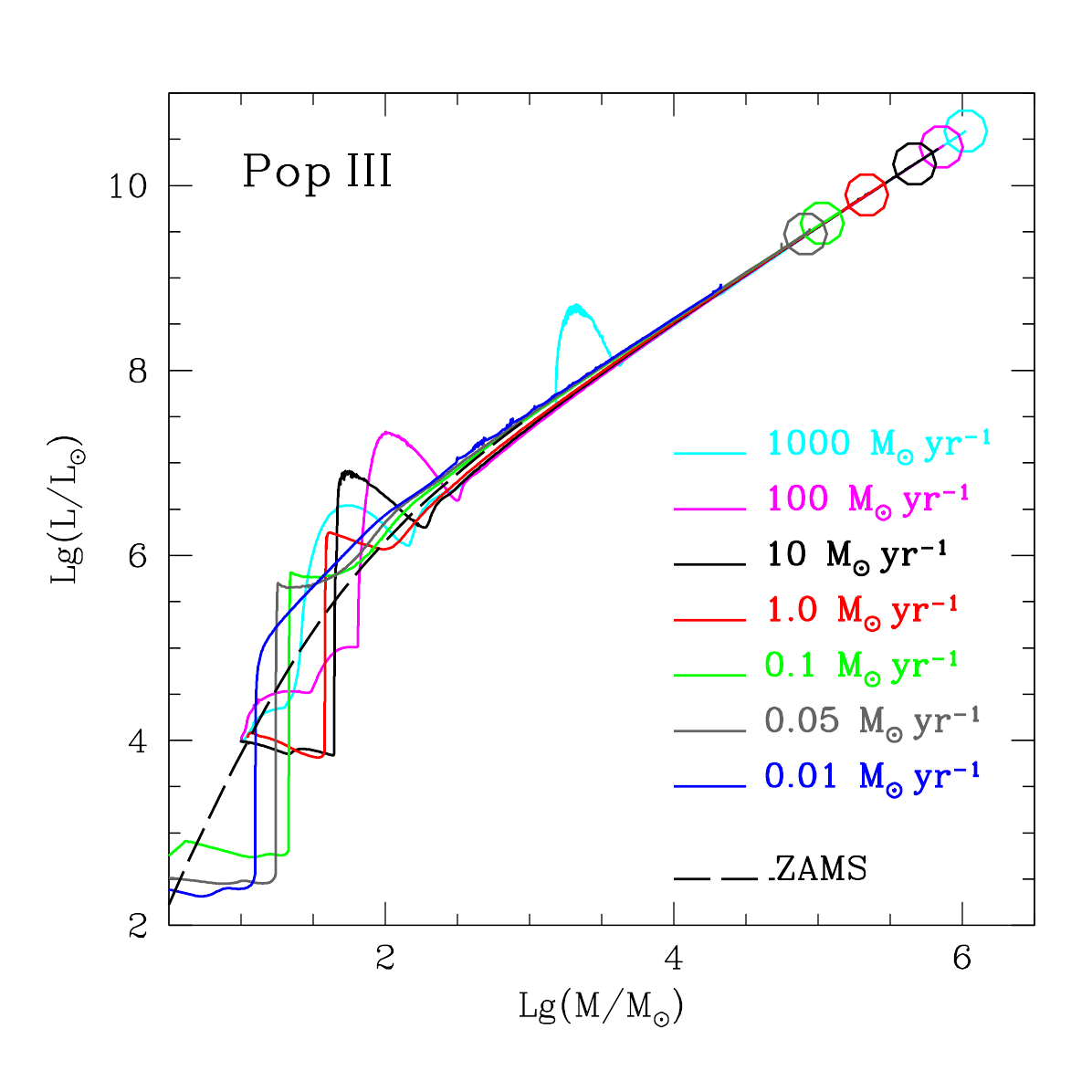} 
    \caption{The Hertzsprung-Russel (HR) diagram (left panel), and the mass-luminosity relation (right panel) for the evolutionary models for accretion rates between $0.01$ and $1000\,\Mpy$. The large open circles indicate the positions where the General relativistic instability occurs, which is discussed in \S\ref{sec:stabil}.     
    The mass-luminosity relationships (right panel) converge towards the Eddington luminosity to mass relation, expressed as $L_{\rm Edd}=4\pi cGM/\kappa_{\rm el}$, where $\kappa_{\rm el}$ denotes the electron-scattering opacity.}
    \label{fig:evolution}
\end{figure*}

Nandal et al.(2024 submitted to A\&A) obtained Pop III SMS evolution models rapidly accreting with a range of  $10^{-6} \le \dot{M}_{\rm acc}[M_\odot{\rm yr}^{-1}] \le 10^3$, using the Geneva stellar evolution code \citep[\gva\,:][]{Eggenberger2008,Lionel2021,Lionel2021b}. 
It is assumed that the accretion of matter occurs via a geometrically thin cold disc which implies that the specific entropy of the accreted matter is the same as that of the stellar surface. Furthermore, any excessive buildup of entropy during accretion is assumed to be radiated away by the surface of the star \citep{Palla1992, Hosokawa2013}. The post-Newtonian gravity is included as per the work of \citet{Lionel2021} which is an extension of the previous work by \cite{Lionel2021b}

Computations of the models begin by accreting primordial matter on hydrostatic seeds. 
 The composition of accreting matter is the same as that of the hydrostatic seed and consists of a homogeneous distribution of hydrogen with a mass fraction $X = 0.7516$, of helium with $Y = 0.2484$ and deuterium at $X(D) = 5 \times 10^{-5}$ \citep{Bernasconi1996, Behrend2001,Lionel2016}.

Among the models computed by Nandal et al.(2024 submitted to A\&A) we have applied our stability analysis to models with accretion rates ranging from $0.01\Mpy$ to $1000\Mpy$.
 Evolution models with $\dot{M}_{\rm acc}\,\ge\,0.01\,\Mpy$ are shown in Fig.\,\ref{fig:evolution} in the HR diagram (left panel) and in the mass-luminosity plane (right panel). Long-dashed lines indicate the relation of the zero-age-main sequence (ZAMS) stars having the same chemical composition. The ZAMS stars, with no accretion energy, are in 
 thermal balance; i.e. the nuclear energy generation rate in the core is exactly balanced with the luminosity at the surface.  
 In accreting stars, accretion heating causes the envelope to expand, resulting in a lower surface temperature compared to a ZAMS model with equivalent luminosity. The lower temperature is limited by the Hayashi limit where a large range of the stellar envelope is in the convective equilibrium. 
 While the surface temperatures (or radii) are very different,
 the luminosity to mass relations of accreting stars are comparable to that of ZAMS.

The mass-luminosity relations shown in Fig.\,\ref{fig:evolution} (right panel) converge to a single relation in the high luminosity range.
The relation corresponds to the Eddington luminosity
\begin{equation}
L_{\rm Edd} = {4\pi cGM\over \kappa_{\rm el}}
=6.5\times10^4{M/M_\odot\over(1+X)}L_\odot,
\label{eq:Led}
\end{equation}
where the electron-scattering opacity is given as $\kappa_{\rm el}=0.2(1+X)$.
The occasional excesses of luminosity above $L_{\rm Edd}$ in relatively low luminosity range can be attributed to the effects of surface convective flux.

Although the luminosity to mass relation in equation (\ref{eq:Led}) has no limiting mass or luminosity, the GR instability occurs at a sufficiently large mass \citep[e.g.,][]{Iben1963,chandrasekhar1964,Osaki1966}.
In this paper, we obtain the critical mass for each accretion rate by solving the equation of the general relativistic adiabatic linear radial pulsation derived by \cite{chandrasekhar1964}. We discuss the method to solve the differential equation in the next section.

%%----------------------------------------------------
\section{GR equations for 
linear adiabatic radial pulsation  
\label{puls} }

We examine the stability of an SMS model by solving the linear pulsation equation 
derived by \citet{chandrasekhar1964}.
Taking into account the GR effects, \citet{chandrasekhar1964} has derived the differential equation for infinitesimal radial displacement $\xi e^{i\sigma ct}$ as

 \begin{equation}  \label{eq:ch64}
 \begin{split}
   { \frac{d}{dr}} \left[e^{3a+b}{\frac{\Gamma_{1}P}{r^2}}{ \frac{d}{dr} }(e^{-a}r^2\xi)\right]= 
    e^{2a+b}\xi\left[ {\frac{4}{r}}{\frac{dP}{dr}}+  {\frac{8\pi G}{c^4}}e^{2b}P(P+c^2\rho) \right. \\
    \left.
    \qquad - {\frac{1}{ P+c^2\rho}}\left(\frac{dP}{dr} \right)^2  
    -\sigma^2 e^{2(b-a)}(P+c^2\rho) \right],
    \end{split}
 \end{equation}
where 
$P$, $\rho$, and $\Gamma_1$ are, respectively, the pressure, matter density,\footnote{ Correction of internal energy \citep{Nagele2022} is not included.} and adiabatic exponent $(d\ln P/d\ln\rho)_{ad}$.
Furthermore, 
$a$ and $b$ are metric  coefficients, with which line element $ds$  is 
given as
\begin{equation}
ds^2=-e^{2a}c^2dt^2 + e^{2b}dr^2+r^2(d\theta^2+\sin^2\theta d\phi^2).
\label{eq:metric}    
\end{equation}
Since the static equilibrium values are sufficient for $a$ and $b$ in Eq.\,(\ref{eq:ch64}), they are obtained from the Einstein field equation for a  
spherically symmetric hydrostatic equilibrium as \citep{chandrasekhar1964}
\begin{equation}
e^{-2b}=1-{2GM_r\over rc^2}, \quad
{e^{-2b}\over r}{d\over dr}(a+b)={4\pi G\over c^4}(P+\rho c^2).
\label{eq:a_b}
\end{equation}
Because Eq.\,(\ref{eq:metric}) at the stellar surface ($r=R$) should be equal to Schwarzshild's metric, we have a relation
\begin{equation}    
a(R)={1\over2}\ln\left(1-{2GM\over Rc^2}\right).
\end{equation}
Equation\,(\ref{eq:ch64})
is a linear eigenvalue equation with eigenvalue $\sigma^2$.
The stellar structure is unstable if a pulsation mode has a negative eigenvalue; i.e., $\sigma^2 < 0$.

To solve the second-order differential equation we introduce, similarly for the Newtonian radial pulsations, two non-dimensional variables, 

\begin{equation}
    Y_1\equiv \frac{\xi}{r} \quad \mbox{and} \quad Y_2\equiv \frac{\Delta P}{P},
\end{equation}   
where $\Delta P$ is the Lagrangian perturbation of pressure.
Using these variables, Eq.\,(\ref{eq:ch64}) can be separated into the two first-order differential equations

\begin{equation}
 \frac{dY_1}{d\ln r}= -\left(3{\bf-}\frac{da}{d\ln r}\right)Y_1 -\frac{1}{\Gamma_1}Y_2,
 \label{eq:dy1}
\end{equation}
and
\begin{equation}
\begin{split}
 \frac{dY_{2}}{d\ln r} =
\left[4\overline{V}-{\bf 2}{\frac{d(a+b)}{d\ln r}} + \overline{V}{\frac{da}{d\ln r}}+
 \omega^2 W e^{2(b-a)} 
 \right] Y_1 \\
\qquad \qquad\qquad
 +\left[\overline{V}-{\frac{d(2a+b)}{d\ln r}} \right]Y_2,
 \label{eq:dy2}
 \end{split}
\end{equation}
where
\begin{equation}
\overline{V}\equiv -{d\ln P\over d\ln r} \quad{\rm and}\quad
 \omega^2 W \equiv \sigma^2r^2 \left(1+{c^2\rho\over P}\right)   
\end{equation}
with the square of non-dimensional pulsation frequency 
\begin{equation}
\omega^2\equiv \sigma^2c^2\left({GM\over R^3}\right)^{-1}.     
\end{equation}
The boundary conditions of being finite at the center for $Y_1$ and at the surface for $Y_2$ are imposed. In addition, we adopt a normalization, $Y_1(R)=1$ at the surface.
The linear homogeneous differential equations (\ref{eq:dy1}) and (\ref{eq:dy2}) with an eigenvalue $\omega^2$ can be solved by the Henyey-type  relaxation method as often used in the analysis for Newtonian stellar pulsations \citep[eg.,][]{Unno1989}. 

We have obtained $\omega^2$ for several stellar structures with accretion rates ranging from 0.01 to 1000 \Mpy.
For a stellar model, we obtain a series of $\omega^2$ depending on the number of nodes in $\xi(r)$. The fundamental mode has no node in $\xi$ and has the smallest $\omega^2$. 
The sign of $\omega^2$ of the fundamental mode determines the stability of the structure. If $\omega^2<0$, we have pure imaginary frequencies $\pm i|\omega$|, so that the GR instability occurs with a growth rate given by $|\omega|\sqrt{GM/R^3}$ 
 in physical dimension. If $\omega^2$ of the fundamental mode is positive, the structure is stable, because perturbations causes only small amplitude pulsations given as the real part of $[\xi(r)\exp(i\omega\sqrt{GM/R^3} t)]$. In this case
$\omega\sqrt{GM/R^3}$ represents an angular frequency of adiabatic radial pulsation.

%%-------------------------------------------------------
\section{Results of the stability analysis}
\label{sec:stabil}
\begin{figure}
    \centering
    \includegraphics[width=0.48\textwidth]{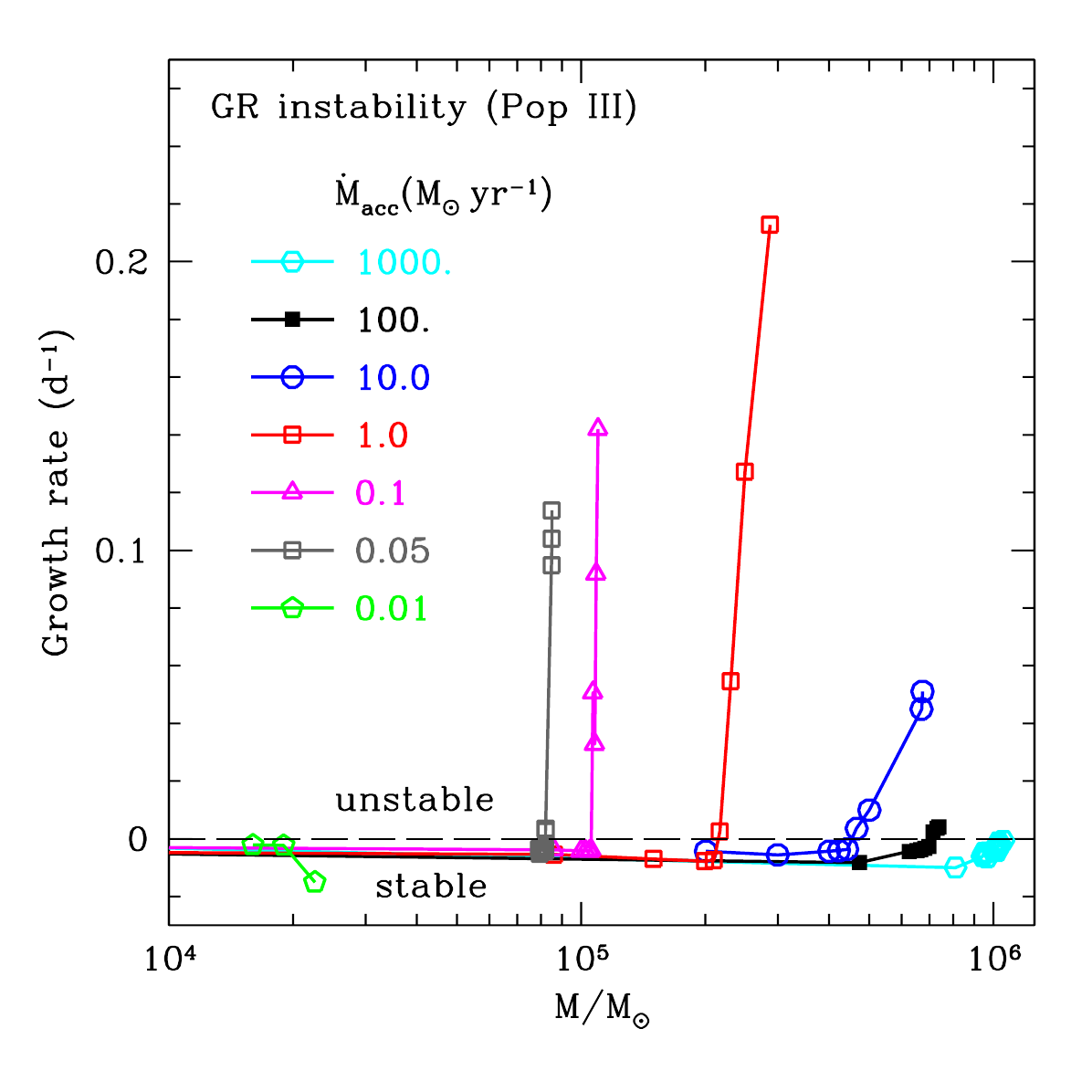}     
    \caption{
    Growth rate of the GRI versus stellar mass for each model.
    The negative part of the vertical axis corresponds to the angular frequency of the stable fundamental pulsation mode.
    The horizontal dashed line indicates the boundary between stable and unstable regions.
        A steep increase in the growth rate indicates that a strong GRI occurs as soon as the mass exceeds the critical mass depending on the accretion rate.
}
\label{fig:growth}
\end{figure}

Figure~\ref{fig:growth} shows the stability of the fundamental modes in selected models transiting from stable to unstable structures (while models with $\dot M_{\rm acc}=0.01\,\Mpy$ remain stable). For an accretion rate of $0.05\,\Mpy$ or higher, the fundamental mode becomes unstable when the mass becomes sufficiently large, which is recognized by the steep increase of the growth rate from a negative value. The critical mass of the GR instability is larger in the models with larger accretion rate, while the increase of the instability growth rate is steeper in the cases of smaller accretion rates (and hence smaller masses). 

The stable side of Fig.\,\ref{fig:growth} shows that the angular frequency $\omega\sqrt{GM/R^3}$ is nearly constant in a wide range of stellar mass.
We can understand this from the model  properties seen in Fig.\,\ref{fig:evolution},
\begin{equation*}
L\approx L_{\rm Edd} \propto M
\quad {\rm and} \quad
L\propto R^2T_{\rm eff}^4 \propto R^2.
\end{equation*}
The second relation is the Stefan-Boltzmann law with an assumption of $T_{\rm eff}\sim$constant.
Combining these relations and using $\omega\sim1$ for the adiabatic fundamental mode, we obtain
\begin{equation*}
\omega\sqrt{GM/R^3} \propto M^{-{1\over4}},  
\end{equation*}
which indicates that the adiabatic pulsation frequency has only a weak dependence on the mass. (The explanation would be improved if we took 
into account the slight luminosity dependence of $T_{\rm eff}$.)

%% -----------------------------------------
\begin{table}[]
    \caption{The Mass, luminosity, and the central helium abundance at GRI }
    %\centering
    \begin{tabular}{ccccc}
  %  \hline
    $\dot{M}$ & $M_{\rm GRI}$ & $\log{L\over L_\odot}$ & $X_{\rm c}$ & ${2GM\over c^2R}$ \\
    $M_\odot{\rm yr}^{-1}$ & $ 10^5M_\odot$ & 
    & & ${\bf10^{-5}}$\\
    \hline
    \\
       0.01  &  $-$  & 8.98$^a$ & 0.00$^a$ & 0.5$^a$\\ 
       0.05  & 0.82 & 9.47 & 0.04 & 1.1 \\
       0.1   & 1.06 & 9.59 & 0.23 & 1.3 \\
        1    & 2.16 & 9.89 & 0.51 & 2.9 \\
       10    & 4.56 & 10.22 & 0.60 & 3.9 \\ 
       100   & 7.08 & 10.41 & 0.73 & 7.1 \\
       1000  & 10.6 & 10.59 & 0.75 & 17. \\ \\
       \hline
       $^a$  final model
    \end{tabular}
    \label{tab:GRI}
\end{table}
% ------------------------------------------
% ------------------------------------------
%% -----------------------------------------

Table\,\ref{tab:GRI} lists obtained critical mass and luminosity as well as the central hydrogen 
abundance ({\bf $X_c$}) at the occurrence of the GRI for each accretion rate.
(For the accretion rate $0.01\Mpy$, quantities of the final model are shown because we have never obtained the GRI in this case.) 
For the models with $0.05\Mpy\lesssim\dot{M}_{\rm acc}\lesssim1000\Mpy$ the GRI occurs during the core hydrogen burning stage, while in the cases of $\dot{M}_{\rm acc}>1000\Mpy$ the GRI would occur before the beginning of the core hydrogen burning.
In contrast, hydrogen at the center of the models with  $\dot{M}_{\rm acc} \lesssim 10^{-2}\,\Mpy$ is exhausted before the mass is increased enough for the GRI. While we did not pursue further evolution, \citet{Nagele2022} found the GRI to occur in (non-accreting) stars of $2\sim3\times10^4\msun$ during or after the core-helium burning stage. 

\begin{figure}
    \centering
\includegraphics[width=0.48\textwidth]{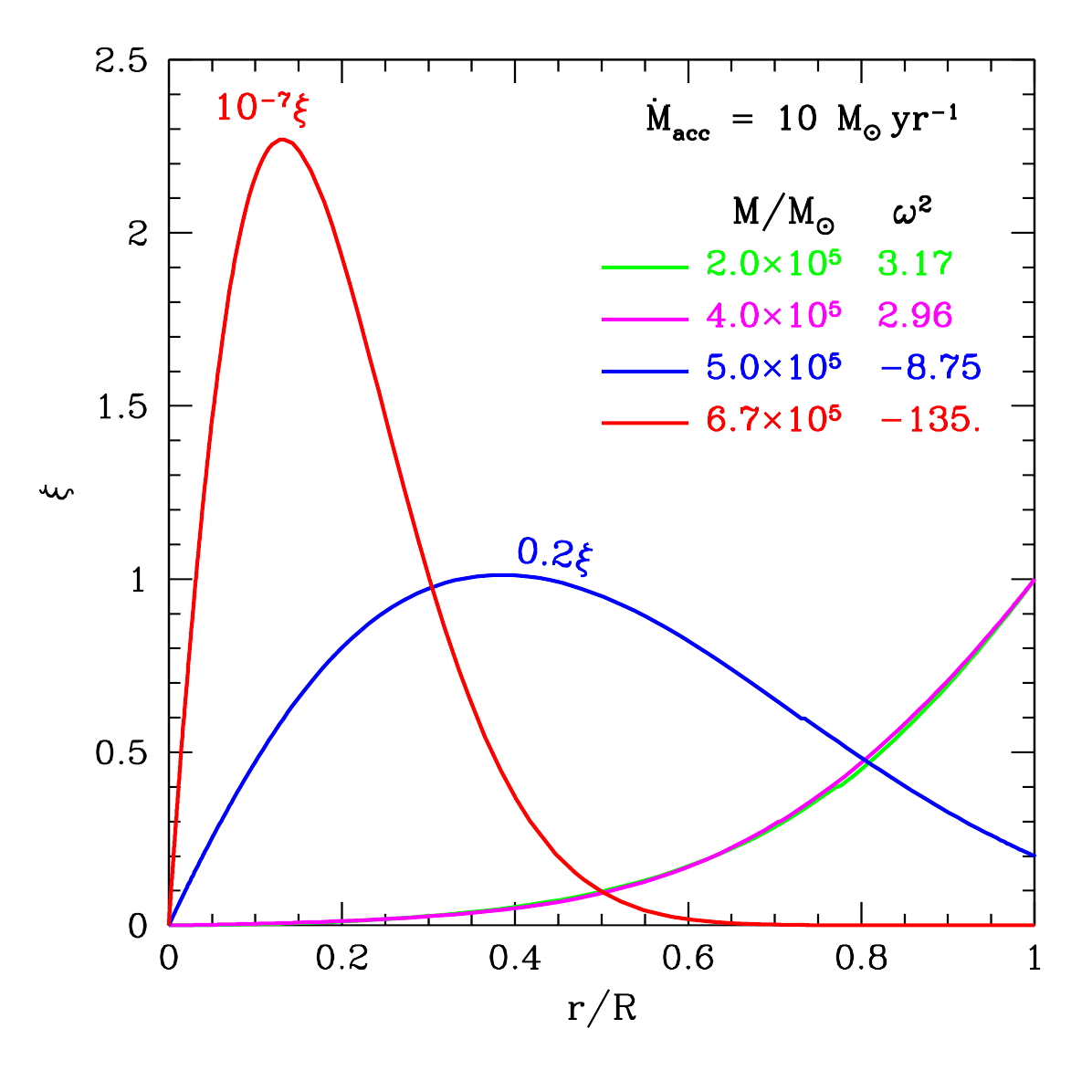}
 \caption{
The displacement $\xi$ of the radial fundamental mode as a function of the distance from the center in some models with an accretion rate of $10M_\odot/{\rm yr}$.
    The displacement is normalised to $\xi=1$ at the surface. For unstable modes ($\omega^2<0$), $\xi$ is scaled by a factor to fit it in the figure.}
    \label{fig:eigfun}
\end{figure}

Figure~\ref{fig:eigfun} shows radial displacements $\xi$ (normalized as unity at the surface) of the fundamental mode as a function of the normalized distance from the center, $r/R$, for models of different masses accreting at $10\,\Mpy$.
Models with $M\lesssim4\times10^5M_\odot$ are stable and the displacement $\xi$ of the fundamental mode is roughly proportional to $r$; i.e. $\xi/r \approx$ constant 
 in outer layers irrespective to the stellar mass.
Models with $M\gtrsim4.5\times10^5M_\odot$ are unstable and the displacement $\xi$ of the fundamental mode ($\omega^2<0$) has a large peak in the interior where the GR effect is strongest. The radial distribution of $\xi$ indicates that the GR instability leads to a collapse of the central part first.
We note, however, that in the mass-coordinate, the $\xi$ peaks are located at $M_r/M \approx 0.92$ and $0.98$, respectively, for the models of $M/\msun=6.7\times10^5$ and $5.0\times10^5$, indicating the GRI to involve most of the mass in the SMS. 

\begin{figure*}
    \centering
    \includegraphics[width=0.48\textwidth]{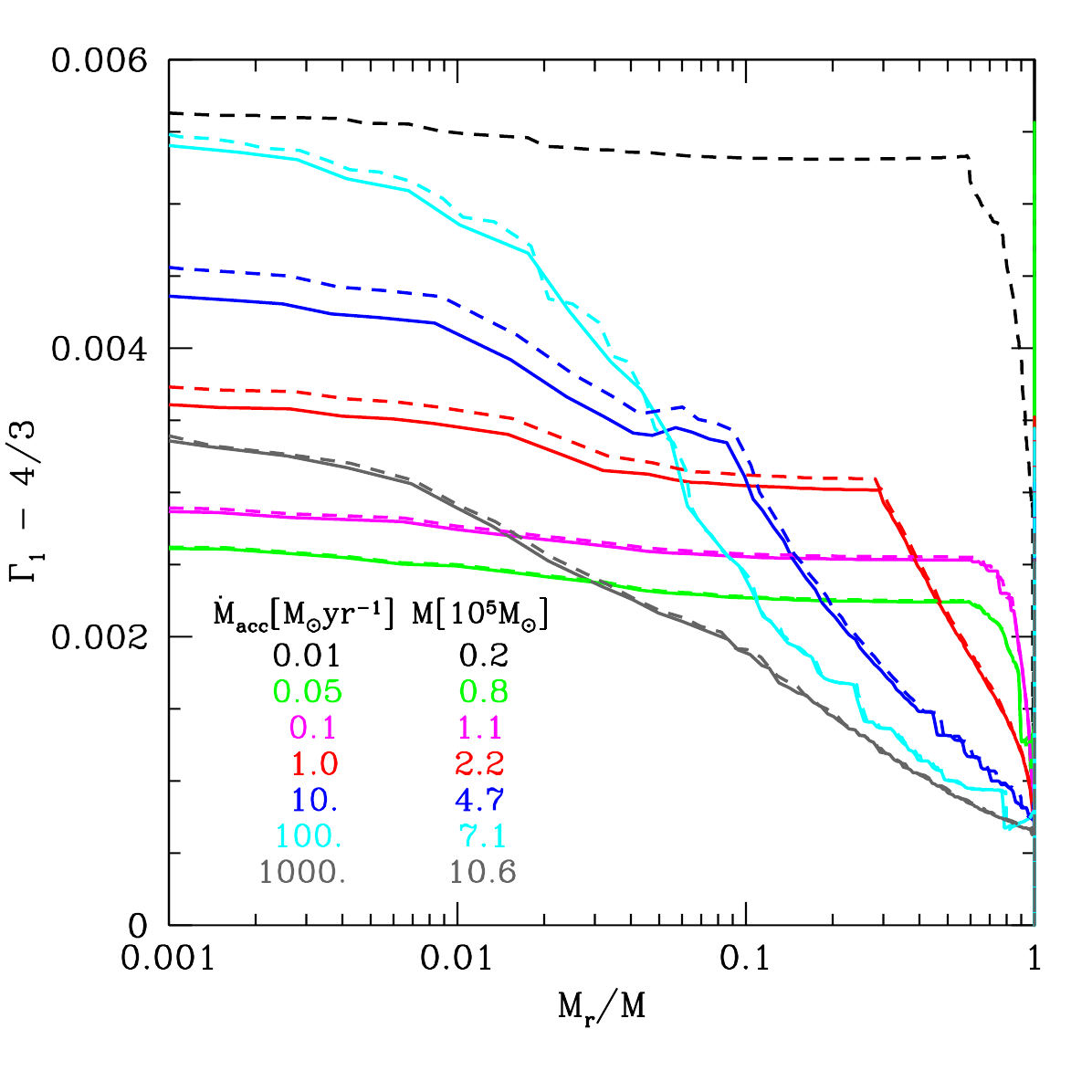}
    \includegraphics[width=0.48\textwidth]{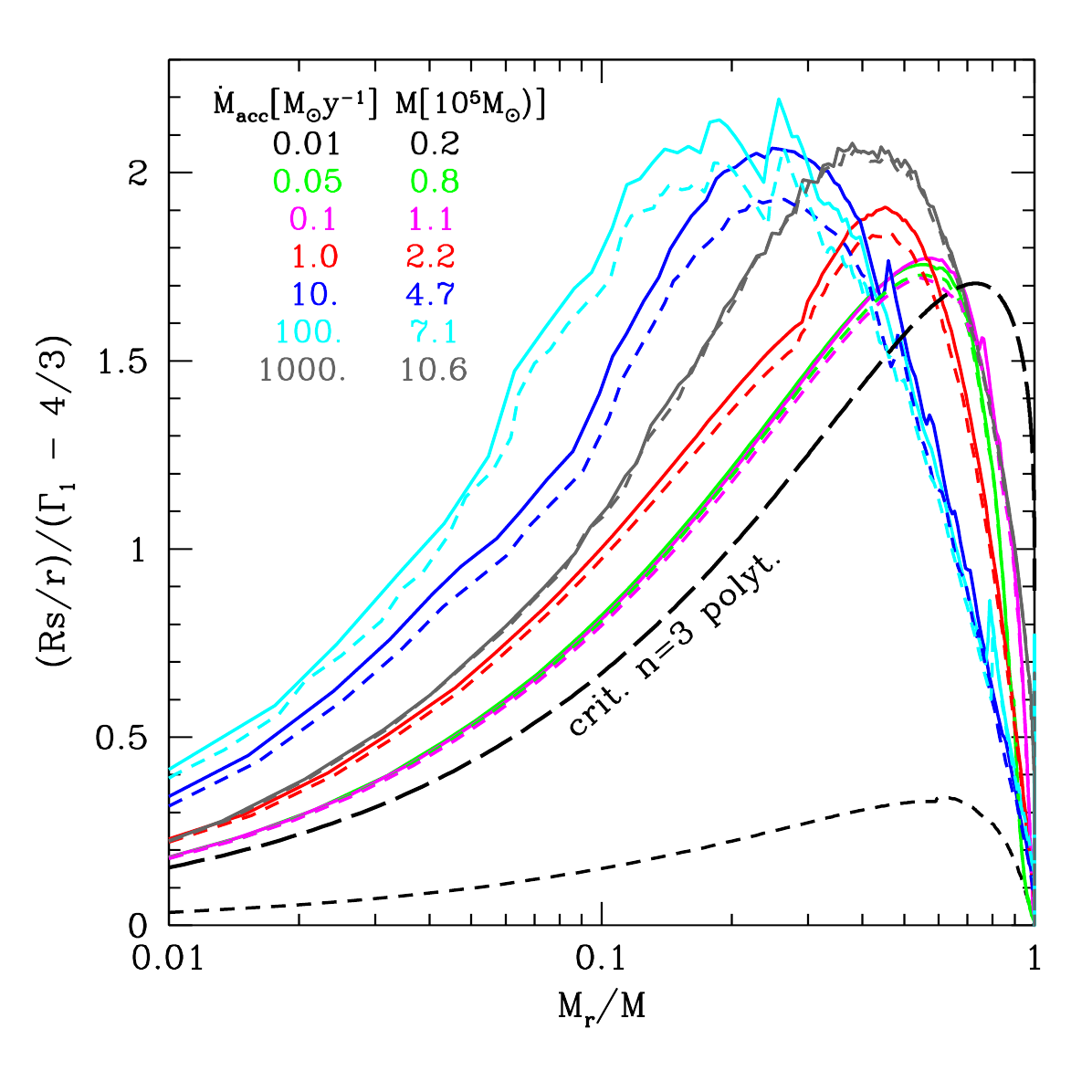}
    \caption{
    {\bf Left:} The difference of the adiabatic exponent $\Gamma_1$ from $4/3$ as a function of $M_r/M$ for selected models. Solid and dashed lines indicate models unstable and stable model against GRI, respectively.   Accretion rates are color coded as indicated. 
    {\bf Right:}
    The quantity $Q$ defined in Eq.\,(\ref{eq:ratio}) is plotted as a function of $M_r/M$ for the structures of SMSs with various accretion rates.  Unstable and stable structures are plotted with solid and short-dashed lines, respectively.
    These curves are compared with the black long-dashed line representing $Q_{\rm crit}$ (Eq.\,\ref{eq:critPol}) that corresponds to the $Q$ for the critical n=3 polytrope. 
    }
    \label{fig:Gam1}
\end{figure*}

\citet{chandrasekhar1964} obtained the condition for the occurrence of the GRI in a polytrope of $n=3$ (under the post Newtonian, PN, approximation) as
\begin{equation}
    \Gamma_1-{4\over3} < 1.1245 {2GM\over c^2 R}.
\label{eq:pol}
\end{equation}
 We have confirmed this relation numerically by applying our stability analysis (\S\ref{puls}) to n=3 PN polytropes for various values of $M/R$ (see Appendix).
However, we cannot directly compare inequality (\ref{eq:pol}) with the occurrence of the GRI in SMS models, because $\Gamma_1$ varies in the interiors of SMSs as shown in the left panel of Fig.\,\ref{fig:Gam1}.
Furthermore, because the outer layers of an accreting SMS model are extended compared with that of the n=3 polytrope,  $2GM/(c^2R)$ at the surface of a GR unstable SMS (Table\,\ref{tab:GRI}) is too small to compare with the inequality (\ref{eq:pol}). 

Taking into account the variable $\Gamma_1$, we consider
the quantity $Q$ given as
\begin{equation}
Q \equiv %{2GM_r\over c^2r}{1\over (\Gamma_1-4/3)}
{R_{\rm S}\over r}{1\over (\Gamma_1-4/3)}
\quad {\rm with} \quad R_{\rm S}\equiv {2GM_r\over c^2}.
\label{eq:ratio}
\end{equation}
The right panel of Fig.\,\ref{fig:Gam1} shows $Q$ as a function of the normalized mass coordinate $M_r/M$
for selected GR unstable (solid lines) and stable models (short dashed line), 
where different accretion rates are color-coded as indicated.
As the evolution proceeds with an accretion rate,
$Q(M_r/M)$ increases as the total mass increases. When the mass becomes 
sufficiently large (depending on the accretion rate), the GRI occurs so that the curve of $Q(M_r/M)$ is switched to a solid line in Fig.\,\ref{fig:Gam1}. 
The stable-unstable transitions can be seen by very closely separated dashed and solid lines. 
It is interesting to note that $Q$ of the unstable model (solid line) is always larger than that of stable model (dashed line) throughout the stellar interior even when the difference is very small, indicating the maximum value of $Q$ (which depends on the accretion rate) indeed determines  the mass at the first encounter with the GR stability.  

For the critical (or neutrally stable) $n=3$ PN polytrope, $Q$ can be written as
\begin{equation}
Q_{\rm crit}={1\over 1.1245}{M_r\over M}{R\over r},
\label{eq:critPol}
\end{equation}
which is derived from Eq.\,(\ref{eq:ratio}) using Eq.\,(\ref{eq:pol}) after replacing the inequality ($<$) with the equality (because of the neutral condition).
The polytropic $Q_{\rm crit}$ is shown by a black long-dashed line   in the right panel of Fig.\,\ref{fig:Gam1}, for comparison with $Q(M_r/M)$ of accreting SMS models.
It is interesting to note that the maximum of $Q_{\rm crit}$ $(\approx\,1.7)$ of the n=3 polytrope to be the lower bound for the $\max(Q)$ of the GR unstable accreting SMSs. 
In other words, the polytropic relation provides the necessary condition, $\max(Q) > 1.7$ for accreting SMS models to be GR unstable. 
The $\max(Q)$ of the unstable less massive SMS for $\dot{M}_{\rm acc}=0.05 - 0.1\,\Mpy$, is comparable with $\max(Q_{\rm crit})$, while the $\max(Q)$ of a more massive SMS with a higher accretion rate ($\dot{M}_{\rm acc} \ge 1\,\Mpy$) is considerably higher than the polytropic limit.
This property is consistent with the fact (left panel of Fig.\,\ref{fig:Gam1}) that $\Gamma_1-4/3$ is nearly constant throughout the interior in the less massive SMS (i.e., structure is comparable with a polytrope), while varies considerably in more massive SMSs.
We note, in passing, that Fig.\,{\ref{fig:Gam1}} clearly indicates that the $2\times10^4M_\odot$ model for the accretion rate $0.01\,\Mpy$ does not satisfy the necessary condition of the GRI.

%% --------------------------------
\section{Comparison with previous results}
\label{sec:comparison}

\defcitealias{Lionel2018}{H18}
\defcitealias{Lionel2021c}{H21b}
\defcitealias{woods2017}{W17}
\defcitealias{umeda2016}{U16}

\begin{figure}
   % \centering
    \includegraphics[width=0.48\textwidth]{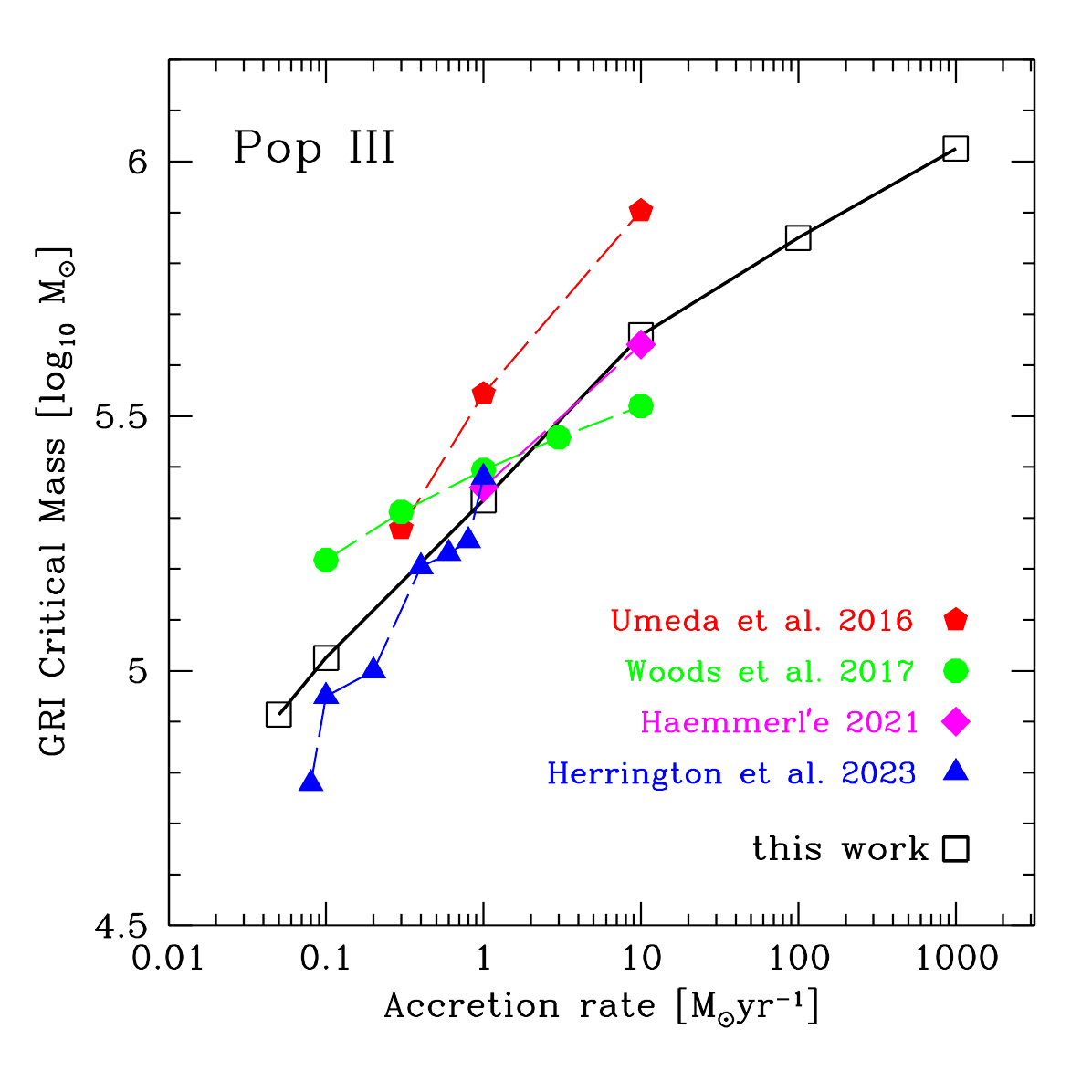}
 \caption{Critical masses of GRI for models with various accretion rates (horizontal axis) are compared with results in the literature. For \citet{Herrington2023}, results from hydrodynamics models (read from their Fig.11) are shown.}
 \label{fig:GRIM}
\end{figure}

Figure\,\ref{fig:GRIM} shows our GRI-critical masses (open squares) versus accretion rates in comparison with the results of previous works.
\citet{umeda2016} determined the GRI critical mass as the mass when the central temperature has increased to $10^{9.2}$K due to hydrostatic contraction. Their masses are systematically larger than those of our models, indicating that GRI collapse should occur earlier if the hydrodynamic effects included. 
This is consistent with the fact that the critical masses from {\it hydrodynamic} models obtained by \citet{Herrington2023} (triangles) are smaller than those obtained from their {\it hydrostatic} models
(not shown in Fig.\,\ref{fig:GRIM}), which are comparable with those of \citet{umeda2016}. On the other hand, the critical masses based on the  hydrodynamic calculations of \citet{Herrington2023} (triangles in Fig.\,\ref{fig:GRIM}) are comparable with our critical masses.

Filled green circles in Fig.\,\ref{fig:GRIM} show the GRI critical masses obtained by \citet{woods2017} using another hydrodynamic evolution code with post Newtonian corrections.
Despite the use of a hydrodynamic code, the green dots (for $\dot{M}< 1\Mpy$ in particular) deviate from the relation of \citet{Herrington2023} (blue filled triangles) as well as from our relation (open squares).
The cause of the deviation is not clear.

The results of \citet{Lionel2021} for the models with
accretion rates of $1\,\Mpy$ and $10\,\Mpy$ (diamond symbols in Fig.\,\ref{fig:GRIM}) agree well with ours. It is reasonable because his stability analysis is also based on Eq.\,(\ref{eq:ch64}) in \S\ref{puls} under the assumption of $\xi/r=$constant, which is a reasonable approximation before the occurrence of the GR instability as seen in Fig.\,\ref{fig:eigfun}.
In addition, his evolution models were obtained by \gva\,  code as Nandal et al. (2024,submitted to A\&A) of which models we are using in this paper.  

%%--------------------------------------
\section{Conclusions \label{sec-conclu}}
We have examined when the GR instability occurs in a rapidly accreting Pop III SMS by solving the linear adiabatic radial pulsation equation derived by \citet{chandrasekhar1964}.
We found the displacement $\xi$ of the fundamental mode of pulsation monotonically 
increases from the center to the surface in a stable model, while it has a large peak between the center and the surface in an unstable model.
The former property of $\xi$ supports the method to find the stable-unstable critical model by assuming $\xi/r$ to be constant \citep{Lionel2021}.

For each accretion rate ($\gtrsim 0.05 \msun {\rm yr}^{-1}$) there is a critical mass above which the GRI occurs and the star begins to collapse. 
The critical mass for an accreting SMS is larger for a larger accretion rate. We found, $8\times10^4\msun$ (at the end of the core hydrogen burning) for $\dot{M}=0.05\Mpy$ and $10^6\msun$ (at the beginning of the core hydrogen burning) for $1000\Mpy$. 

Comparisons with previous results in the literature show that our results agree well with \citet{Lionel2021}'s results for $1$ and $10\Mpy$, and  hydrodynamic calculations of \citet{Herrington2023} for $0.1$ to $1\Mpy$. Our present work has extended critical masses for the GRI from those of the previous works to $7\times10^5$ and $10^6\msun$ for the accretion rates 100 and $1000\Mpy$, respectively.

\begin{acknowledgements}
We are grateful to the referee, Dr. Chris Nagele for very useful comments and suggestions, which have greatly improved this paper.
The authors of this paper have received supports
from the European Research Council (ERC) under the European Union's Horizon 2020 research and innovation program (grant agreement No 833925, project STAREX).
\end{acknowledgements}

\bibliographystyle{aa}
\bibliography{bibliography}

\begin{appendix}
 \section{Stability analysis for Post-Newtonian Polytrope}   
To test our radial pulsation code, we have applied it to an n=3  polytrope under the Post-Newtonian approximation for which the critical condition for the GRI has been derived by \citet{chandrasekhar1964}. 
In this appendix we summarize equations needed in the adiabatic radial pulsation analysis for a polytrope under the Post-Newtonian approximation. We refer the reader  \citet{chandrasekhar1964} for details.

\subsection{Equilibrium Post-Newtonian Polytrope}
The pressure $P$ and the density $\rho$ are expressed as
\begin{equation}
    P=P_{\rm c}\Theta^{n+1} \quad {\rm and} \quad 
    \rho = \rho_{\rm c}\Theta^n,
\end{equation}
where the subscript 'c' indicates the value at the center, and $n$ is the polytropic index. In the Post-Newtonian (PN) approximation, $\Theta$ is expressed as
\begin{equation}
    \Theta = \theta + q\phi  \quad {\rm with} \quad
    q= {P_{\rm c}\over\rho_{\rm c}c^2}={\eta_1\over 2(n+1)v_1}{2GM\over Rc^2},
\end{equation}
where $\theta$ is the Lane-Emden function of $\eta$; dimension-less distance from the center defined as
\begin{equation}
    \eta = r/\alpha  \quad {\rm with} \quad
    \alpha\equiv \left[{(n+1)qc^2\over 4\pi G\rho_{\rm c}}\right]^{1/2}.
\end{equation}
The subscript '$_1$' indicates the value at first zero of $\theta$.
The classical Lane-Emden function $\theta$ and an additional function $\phi$ representing the PN effects satisfy the following equations:
\begin{equation}
{1\over\eta^2}{dv\over d\eta} = \theta^n \quad {\rm with} \quad
v=-\eta^2{d\theta\over d\eta}
\end{equation}
and
\begin{equation}
  \eta^2{d\phi\over d\eta} = -\left[\theta+2(n+1){v\over\eta}\right]v-\eta^3\theta^{n+1}-w 
  \quad {\rm with} \quad
  {dw\over d\eta}=n\eta^2\theta^{n-1}\phi.
\end{equation}
We have obtained functions $\theta(\eta)$, $v(\eta)$, $\phi(\eta)$ and $w(\eta)$ by integrating above differential equations from a point very near the center to the first zero poing of $\theta$.
For the initial values at the starting point where $\eta\ll 1$, we have used the following relations:
\begin{equation}
    \theta\approx 1-{\eta\over6}, \quad \phi \approx -{2\over3}\eta^2,\quad {\rm and} \quad w\approx -{2n\over15}\eta^5.
\end{equation}

\subsection{Adiabatic radial pulsation analysis}
Having polytropic functions, $\theta$,$v$,$\phi$, and $w$, we can evaluate metrics $a$ and $b$ as
\begin{equation}
    e^{-2b} =1 - 2(n+1)q{v\over\eta} \quad {\rm and} \quad
    e^{2a} = 1 - 2(n+1)q\left(\theta+{v_1\over\eta_1}\right).
\end{equation}
Using these relations we express quantities needed in the differential equations (Eqs.\,\ref{eq:dy1} and \ref{eq:dy2}) for $Y_1$ and $Y_2$ as
\begin{equation}
    {da\over d\ln\eta} = (n+1)q{v\over\eta} \quad {\rm and} \quad
    {d(a+b)\over d\ln\eta} =(n+1)q\eta^2\theta^n,
\end{equation}
\begin{equation}
    \overline{V}= {(n+1)\over\theta+q\phi}\left[{v\over\eta}+q\left(\theta+2(n+1){v\over\eta}\right){v\over\eta}+q\eta^2\theta^{n+1}+q{w\over\eta}\right]
\end{equation}
and
\begin{equation}
    We^{2(b-a)}={(n+1)v_1\eta^2(q+\theta^{-1})\over \eta_1^3[1-2(n+1)q(\theta+v/\eta+v_1/\eta_1)]}.
\end{equation}

\begin{figure}
    \centering
    \includegraphics[width=0.48\textwidth]{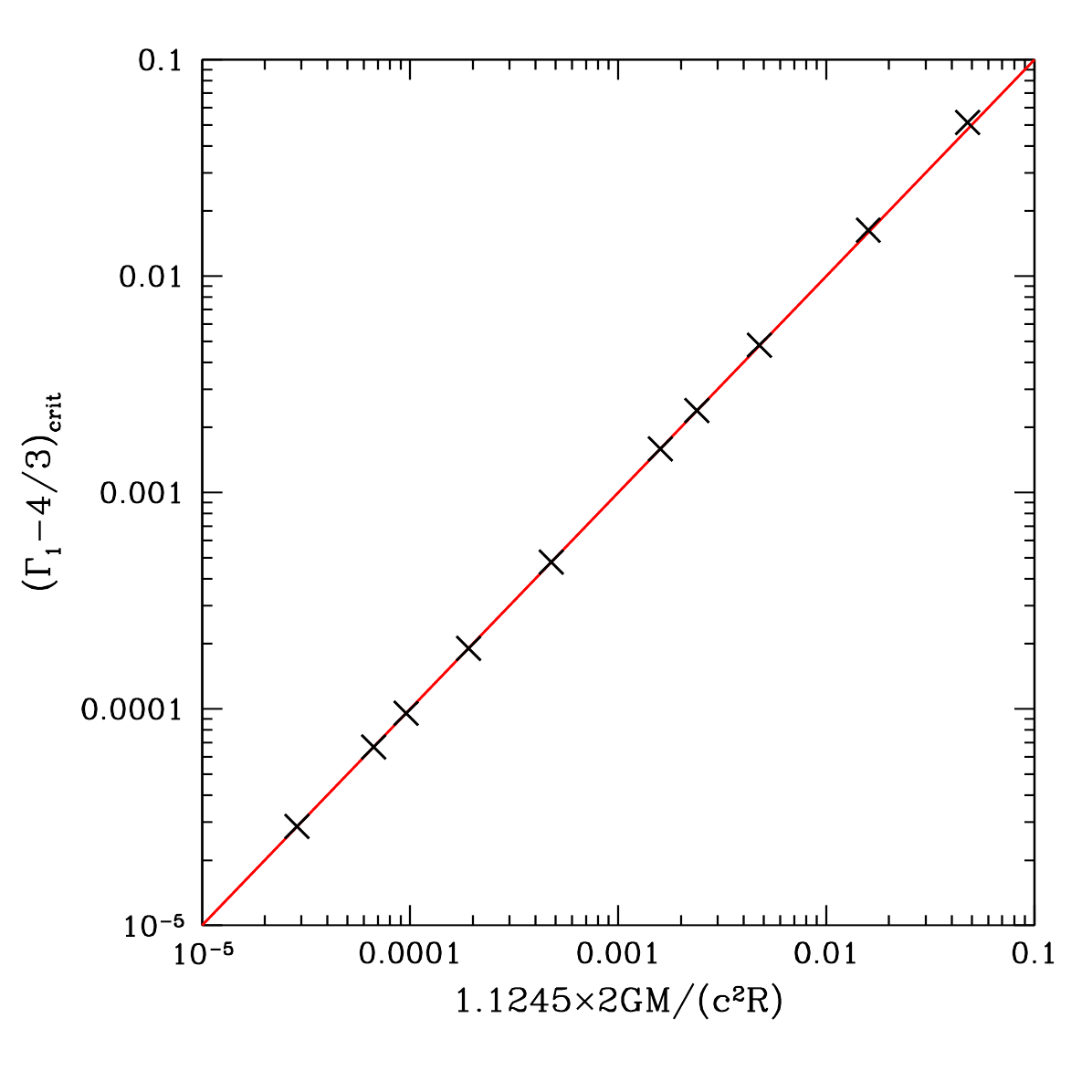}
    \caption{Numerically determined critical $\Gamma_1-{4\over3}$ for the GRI for Post Newtonian n=3 polytropes versus analytical prediction $1.1245{2GM\over c^2R}$.}
    \label{fig:poly3}
\end{figure}

For the numerical test we have adopted $n=3$.
For each assumed values of $M/R$'s we determined the critical value of $(\Gamma_1-{4\over3})$ for the neutral stability, and compared with the theoretical limit $1.1245\times 2GM/(c^2R)$.
As shown in Fig.\,\ref{fig:poly3}, we have obtained a very good agreement, which give us confidence in the discussions presented in the main text of the present paper.

\end{appendix}
\end{document}